\documentclass[prd,aps,twocolumn,nofootinbib]{revtex4}
\usepackage{graphicx} 

\usepackage{color}
\def\gr{$\gamma$-ray}
\begin{document}

\title{Neutrino emission and corona heating induced by high-energy proton interactions in Seyfert galaxies}
\author{A.~Neronov$^{1,2}$}
\author{O.~Kalashev$^{2}$}
\author{D.~V.~Semikoz$^{1}$}
\author{D.~Savchenko$^{1,3,4}$}
\author{M.~Poleshchuk$^{3}$}

\affiliation{$^{1}$Universit\'e Paris Cit\'e, CNRS, Astroparticule et Cosmologie, 
F-75013 Paris, France}

\affiliation{$^{2}$Laboratory of Astrophysics, \'Ecole Polytechnique F\'ed\'erale de Lausanne, CH-1015 Lausanne, Switzerland}

\affiliation{$^{3}$Bogolyubov Institute for Theoretical Physics of the NAS of Ukraine, 03143 Kyiv, Ukraine}

\affiliation{$^{4}$Kyiv Academic University, 03142 Kyiv, Ukraine}

\begin{abstract}
    Recent detection of very-high-energy neutrino emission from Seyfert type active galactic nuclei (AGN) provides a new insight into the physics of the AGN central engines. We notice that if high-energy protons responsible for neutrino emission are accelerated close to the surface of the accretion disk, the neutrino flux may have no unambiguously identifiable electromagnetic counterpart. This is because the electromagnetic power released in interactions of high-energy protons would only contribute to the heating of the disk surface and corona above the disk, rather than escape from the source. Given that the heat deposited in the corona is released in the hard X-ray range we notice that there still might be an ``indirect'' electromagnetic counterpart of the neutrino signal: the hard X-ray flux variability may be strongly or weakly correlated with the neutrino flux variations, depending on the importance of the high-energy proton heating in the disk/corona heat balance. If heating by high-energy protons provides a sizable contribution to the overall corona heating rate, the overall flux of diffuse GeV neutrino background from Seyfert galaxies may be comparable to the X-ray background flux and the high-energy tail of this background can  provide a sizable contribution to the astrophysical neutrino flux in the TeV band. 
\end{abstract}

\maketitle

\section{Introduction}

Evidence for neutrino signal from Seyfert galaxies \cite{IceCube:2022der,Neronov:2023aks,Abbasi:2024hwv,IceCube:2024dou} has revealed an unexpected aspect of high-energy activity of  this otherwise ``high-energy quiet" class of Active Galactic Nuclei (AGN). Neutrinos observed by IceCube telescope can be produced in interactions of high-energy protons and atomic nuclei with either low-energy protons or photons. In both cases, the neutrino production is accompanied by emission of other high-energy particles, including \gr s, electrons, positrons, with the overall power output in the electromagnetic channel comparable to the neutrino power \cite{Murase:2019vdl,Kheirandish:2021wkm,Eichmann:2022lxh,Inoue:2024nap}. 

Gamma-rays produced in decays of neutral pions have spectral characteristics similar to those of neutrinos. If the gamma-rays were able to freely escape from the source, their flux and spectrum would serve as a good proxy for the neutrino flux and spectrum. This is not the case for the individual Seyfert galaxies detectable by IceCube. The gamma-ray luminosity of the sources is much below their neutrino luminosity in the energy band of the neutrino signal \cite{MAGIC:2019fvw,Peretti:2023crf,Murase:2023ccp}.  This suggests that the electromagnetic power  is re-processed by an electromagnetic cascade inside a compact source close to the black hole powering the AGN. Models of such cascade suggest that the "non-thermal" electromagnetic power output in the 1-10 MeV energy range scales with the neutrino power output. Such a correlation  may have escaped detection up to now, because of the absence of (sensitive enough) telescopes in this energy range \cite{Murase:2019vdl,Kheirandish:2021wkm,Eichmann:2022lxh}. 

 Multi-messenger data  are consistent with the possibility that the neutrino luminosity of Seyfert galaxies correlates with the hard X-ray luminosity, once corrected for the attenuation of the hard X-ray flux due to Compton scattering \cite{Neronov:2023aks, Abbasi:2024hwv}. Seyfert galaxies that show evidence for a neutrino signal are the sources with the highest intrinsic hard X-ray flux. The hard X-ray emission from Seyfert galaxies is produced by the corona heated to temperature  $T_c\sim 10^2$~keV, much higher than the temperature in the innermost part of the accretion disk ($T_d\lesssim 10^2$~eV) \cite{1994ApJ...436..599S,1991ApJ...380L..51H,1994ApJ...432L..95H,Fabian:2017ttr}.  The details of the mechanism that maintains high temperature in the AGN corona are not completely known. Uncertainties of the location of the heat source are responsible for large variety of models of geometry of the hot corona ("slab", ``lamppost", ``patchy", ``outflowing" geometries are considered) \cite{2015MNRAS.448..703W,Wilkins:2015ela,Dovciak:2015hda}. It is conventionally assumed that, similar to the Solar corona, a possible source of energy for the heating of the AGN corona is dissipation of the energy of the turbulent magnetic field at the surface of the black hole accretion disk \cite{1998MNRAS.299L..15D,Merloni:2000gs}. The same mechanism should also lead to the acceleration of particles, via reconnection \cite{Guo:2023ybt} and turbulence \cite{Lemoine:2024roa}. Particle acceleration with subsequent dissipation of the energy of the accelerated particles may actually be part of the heating process.

In what follows, we explore this possibility. In this  scenario, the neutrino and hard X-ray luminosities are correlated because the energy source of corona heating is the same as that of particle acceleration, and that heating via particle acceleration may provide a sizable contribution to the overall corona heating rate. 

\section{Neutrino emission from the surface of AGN accretion disk}

High-energy neutrinos can be produced in interactions of high-energy protons (and atomic nuclei) with either low-energy protons or with photons. Both mechanisms should be active in the vicinity of the supermassive black hole powering an AGN. The black hole accretion disk provides a relatively dense medium, with particle density \cite{Jiang:2019ztr} $n_{p,d}\sim 10^{16}.. 10^{18}\mbox{ cm}^{-3}$
 for the black holes of moderate mass $M_{BH}\lesssim 10^8M_\odot$.

The accretion disk spectrum is conventionally modeled as a multi-temperature black body with the temperature scaling with the distance from the black hole. Close to the black hole, the black body spectrum peaks in the ultraviolet range (temperatures $T_d\sim 30$~eV, photon energies  $E_{ph, d}\sim 3T\sim 100$~eV)   \cite{Abramowicz:2011xu}, with the photon density reaching
\begin{eqnarray}
    &&n_{ph,d}\sim \frac{L_{d}}{4\pi R_{d}^2E_{ph,d}}\simeq \\ &&10^{17}\left[\frac{L_{d}}{10^{45}\mbox{ erg/s}}\right]\left[\frac{R_{d}}{10^{13}\mbox{ cm}}\right]^{-2}\left[\frac{E_{ph,d}}{10^2\mbox{ eV}}\right]^{-1}\mbox{ cm}^{-3}\nonumber
\end{eqnarray}
at the distances close to the innermost stable circular orbit, which can reach $R_{d}=3R_{Schw}$ for a non-rotating black hole with the Schwarzschild radius
$R_{Schw}\simeq 3\times 10^{12}\left[M/10^7M_\odot\right]\mbox{ cm}$
and can be as small as $R_{d}=0.5R_{Schw}$ for a maximally rotating black hole.

Motions of plasma in the disk lead to the generation of a magnetic field. The strength of magnetic field with energy density in equipartition with kinetic energy density of the disk matter can reach
\begin{equation}
    B_{d}\lesssim \sqrt{6\pi n_{p,d} T_d}\sim 10^4\left[\frac{n_{p,d}}{10^{17}\mbox{ cm}^{-3}}\right]^{1/2}\left[\frac{T_d}{30\mbox{ eV}}\right]^{1/2}\mbox{ G}
\end{equation}
Even if the magnetic field energy density is sub-dominant compared to the plasma energy density inside the disk, it can be dominating at its surface at the base of corona, where the density is much lower. In fact, dissipation of the magnetic field energy through reconnection and turbulence is commonly assumed to be the source of heating of the corona \cite{Merloni:2000gs}.

Magnetic reconnection and turbulence are also generically 
expected to lead to particle acceleration. The accelerated particles move along tangled magnetic field lines, mostly anchored in the disk. This means that most of the high-energy particles finally re-enter the disk. 
Such particles dissipate their energy in interactions with disk matter and radiation fields.

Protons with energies lower than the pion production threshold $E_{thr,pp}\simeq 0.2 \mbox{ GeV}$ mostly suffer from non-radiative Coulomb energy loss~\cite{pdg}, with the energy loss distance scale
much shorter than the scale height of the accretion disk ($H_d\lesssim R_d)$.
Higher-energy protons loose energy mostly through pion production reaction on the distance scale 
\begin{equation}
\lambda_{pp}=\frac{1}{\kappa\sigma_{pp}n_{p,d}}\simeq 10^9\left[\frac{n_{p,d}}{10^{17}\mbox{ cm}^{-3}}\right]^{-1}\mbox{ cm,} 
\end{equation}
which is still much shorter than the disk size scales ($\sigma_{pp}\simeq 3\times 10^{-26}$~cm$^2$ is the pion production cross-section, $\kappa\sim 0.5$ is the inelasticity of collisions). In such a situation, the disk works as an efficient "beam dump" for the accelerated protons. The production and decay of charged pions result in efficient neutrino emission from the disk.


An alternative neutrino production channel is via interactions of high-energy protons  with energies above a threshold $
    E_{thr,p\gamma}\simeq 10^{15}\left[E_{ph,d}/10^2\mbox{ eV}\right]^{-1}\mbox{ eV}$
with the photon background in the disk. The cross-section of pion production in $p\gamma$ interactions is lower than that of the $pp$ interactions. Even if the density of protons and soft photons in the disk is comparable, it is the $pp$ interaction channel that is responsible for most of the neutrino production. 

Still one more interaction channel for high-energy protons is electron-positron pair production in interaction with both matter and radiation in the disk. Compared to the photo-pion production, this process has a lower energy threshold, 
$E_{thr,p\gamma}'\simeq 5\times 10^{12}\left[E_{ph,d}/10^2\mbox{ eV}\right]^{-1}\mbox{ eV}$ and larger cross-section, $\sigma_{p\gamma}'\gtrsim 10^{-26}$~cm$^2$, but smaller inelasticity. At equal densities of the disk matter and photon fields, it is still the pion production in proton-proton interactions that dominates the energy-loss rate. Pair production in interactions with low-energy protons is always a sub-dominant energy loss channel. 

If the spectrum of high-energy protons is a powerlaw, the spectrum of neutrinos from proton-proton interactions is also a powerlaw with the slope close to that of the parent proton spectrum. IceCube measurements of the neutrino spectrum directly provide information on the spectrum of protons accelerated in the AGN core in the TeV-PeV energy range, but not at lower energies down to the threshold of pion production in proton-proton interactions (in sub-GeV energy range). There is currently no neutrino telescope sensitive enough in this energy to detect a neutrino signal from cosmic sources on top of an overwhelmingly stronger atmospheric neutrino background. This means that the total power of neutrino emission from Seyfert galaxies is currently not known.

\section{Heating of the disk surface and base of corona by high-energy protons}

Apart from neutrinos, pion decays also produce \gr s, electrons and positrons, collectively forming an "electromagnetic" pion decay channel. The total energy injected into the electromagnetic channel is comparable to that in the neutrino channel. The highest energy electrons loose energy mostly through inverse Compton scattering and synchrotron emission. 
The synchrotron energy loss cooling distance  is 
\begin{equation}
    \lambda_s\simeq 6\times 10^{8}\left[\frac{B_d}{10^4\mbox{ G}}\right]^{-2}\left[\frac{E_e}{10^{8}\mbox{ eV}}\right]^{-1}\mbox{ cm}
     \label{sync_loss}
\end{equation}
where $E_e$ is the electron energy, while the inverse Compton cooling time in Thomson regime is
\begin{equation}
    \lambda_{IC}\simeq 3\times 10^{8}\left[\frac{E_{ph,d}n_{ph,d}}{10^{19}\mbox{ eV/cm}^3}\right]^{-1}\left[\frac{E_e}{10^{8}\mbox{ eV}}\right]^{-1}\mbox{ cm}
\end{equation}
Both synchrotron and inverse Compton cooling distances are much shorter than the size of the system, which means that  electrons release their energy right in the surface layer of the disk, while cooling down to at least 100 MeV energy range.

Lower energy electrons are predominantly cooled by Bremsstrahlung and Coulomb losses that operate on distance scales  
\begin{equation}
    \lambda_{Brems}\simeq 6\times 10^{8}\left[\frac{n_{d}}{10^{17}\mbox{cm}^{-3}}\right]^{-1}\mbox{ cm}
\end{equation}
and 
\begin{equation}
    \lambda_{Coul,e}\simeq 5\times 10^8\left[\frac{n_{d}}{10^{17}\mbox{ cm}^{-3}}\right]^{-1}\left[\frac{E_e}{10^8\mbox{ eV}}\right]\mbox{ cm}
\end{equation}

Synchrotron, inverse Compton and Bremsstrahlung losses support the development of electromagnetic cascade, because they result in production of photons that in their turn also interact. Similar to protons and electrons, the propagation distance of  photons  in the disk is much shorter than the disk dimensions. Photons with energies higher than
$E_{thr,\gamma\gamma}\simeq 10\left[E_{ph,d}/10^2\mbox{ eV}\right]^{-1}\mbox{ GeV}$
are absorbed in pair production on the UV radiation from the  disk. The mean free path with respect to this process  is  
\begin{equation}
\lambda_{\gamma\gamma}=\frac{1}{\sigma_{\gamma\gamma}n_{ph,d}}\simeq 3\times 10^{7}
 \left[\frac{L_{d}}{10^{45}\mbox{ erg/s}}\right]^{-1}\left[\frac{R_{d}}{10^{13}\mbox{ cm}}\right]^{2}\mbox{ cm,}
\end{equation}
(where $\sigma_{\gamma\gamma}\simeq 10^{-25}$~cm$^2$ is the pair production cross-section).
Photons with energies above the pair produciton threshold $E_{thr,p\gamma}\simeq 1\mbox{ MeV}$
also produce electron-positron pairs in the Coulomb field of atomic nuclei. The cross-section of this process grows logarithmically with photon energy in the range $\sigma_{BH}\sim 10^{-27}...10^{-26}$~cm$^2$, so that the mean free path of the $E_\gamma>1$~MeV photons is limited to 
\begin{equation}
\lambda_{BH}=\frac{1}{\sigma_{BH}n_{d}}\simeq 10^9...10^{10}\left[\frac{n_{d}}{10^{17}\mbox{ cm}^{-3}}\right]^{-1}\mbox{ cm}
\end{equation}
weakly depending on the photon energy.
Lower energy photons $E_\gamma\lesssim 1$~MeV are Compton scattered on still shorter distance scale, because of the larger cross-section of the Compton scattering:
\begin{equation}
\lambda_{C}=\frac{1}{\sigma_{T}n_{d}}\simeq 10^7\left[\frac{n_{d}}{10^{17}\mbox{ cm}^{-3}}\right]^{-1}\mbox{ cm}
\end{equation} 

Overall, the distance scales of all reactions involving electrons and photons (electromagnetic particles) of all energies are much shorter than the geometrical dimensions of the disk and all the electromagnetic power is ultimately transformed into heat. 

\begin{figure}
\includegraphics[width=\linewidth]{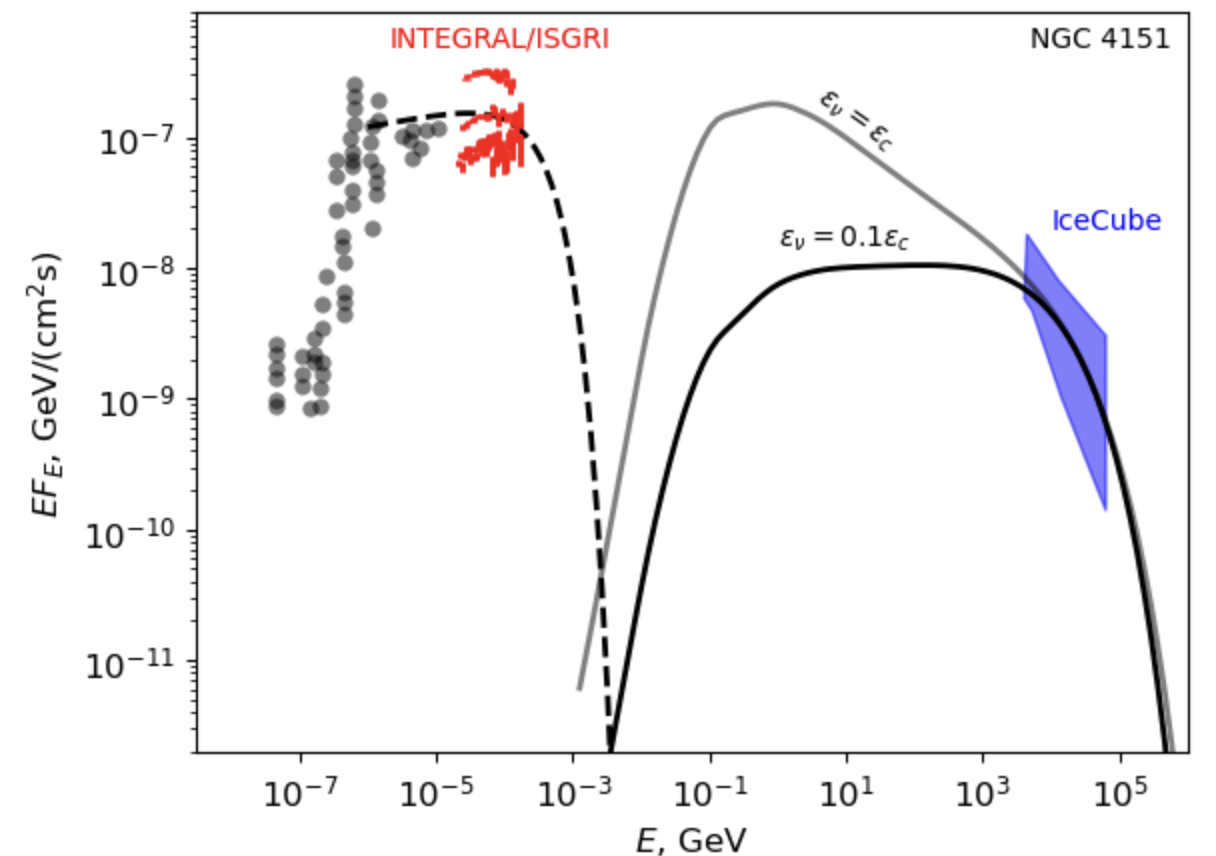}
\includegraphics[width=\linewidth]{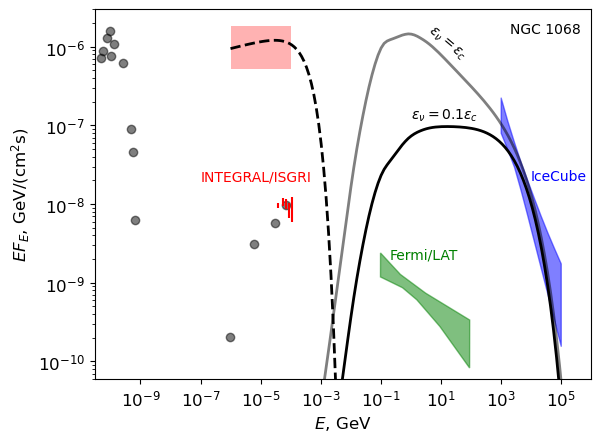}
    \caption{Spectral energy distributions of NGC 4151 (top panel) and NGC 1068  (bottom panel).  Blue butterflies are the all-flavour neutrino spectrum measured by IceCube \cite{Abbasi:2024hwv}, green butterflies are the \gr\ spectra measured by Fermi/LAT (\cite{2010tsra.confE.124L} for NGC 1068). Red hard X-ray data are from INTEGRAL/ISGRI (\cite{2010MNRAS.408.1851L} for NGC 4151,  \cite{2010tsra.confE.124L} for NGC 1068). Grey points show X-ray data from NASA Extragalactic Database (https://ned.ipac.caltech.edu/). Solid lines show model spectra of neutrino emission from interactions of high-energy protons calculated using AAFrag package \cite{Koldobskiy:2021nld}. The neutrino spectra are normalized in such a way that the total power of the neutrino emission is a fixed fraction of the total power of the hot corona,  $\epsilon_\nu=0.1 \epsilon_c$ (black) and $\epsilon_\nu=\epsilon_c$ (gray). 
     Dashed lines show a template spectrum of emission from hot corona (a cut-off powerlaw with the slope $\Gamma=1.9$ and high-energy cut-off at $E_{cut}=300$~keV). 
    }
    \label{fig:spectrum} 
\end{figure}

Heating of the surface of accretion disk by particle showers initiated by high-energy protons is part of the overall process of heating due to dissipation of the energy of magnetic field, that is thought to be responsible for maintaining the temperature of the hot corona above the accretion disk. It is not clear a-priori, what is the fraction $\epsilon_p$ of magnetic energy initially transferred to the accelerated protons, compared to other channels of magnetic energy dissipation. It is also not clear a-priori, which fraction $\epsilon_c$ of the heat absorbed in the surface layer of the disk is transferred to the corona.  If $\epsilon_p$ is large, $\epsilon_p\lesssim 1$, a non-negligible part of the magnetic field energy may be transferred to the high-energy protons and then to neutrinos and electromagnetic cascades in nearly equal amounts ($\epsilon_\nu\sim \epsilon_{em}\sim \epsilon_p/2$). The heat fraction $\epsilon_c$ deposited in the corona and ultimately dissipated via hard X-ray emission from electrons upscattering photons from the accretion disk (Comptonisation process) can in principle be $\epsilon_c\sim \epsilon_p \sim \epsilon_\nu$. Such a possibility is considered in Fig. \ref{fig:spectrum}, showing the spectral energy distributions of NGC 1068 and NGC 4151. The red data points and a red horizontal band show the levels of observed and intrinsic hard X-ray fluxes of the sources. The neutrino flux, assumed to originate from the parent cut-off powerlaw proton distribution, is normalized in such a way that the total neutrino luminosity is equal to the hard X-ray source luminosity ($\epsilon_\nu\sim \epsilon_c$). The high-energy proton spectrum has a slope $\Gamma_p=2.4$  and a high-energy cut-off at the proton energies $E_{p,cut}=500$~TeV (for NGC 4151) or $50$~TeV (for NGC 1068). One can see that such spectral models provide a reasonable fit to the neutrino data and do not violate any observational constraints. Alternatively, assuming a lower ratio of the energy fractions $\epsilon_p\sim 0.1\epsilon_c$, a harder neutrino spectrum with $\Gamma=2$ can fit the data, with high-energy cut-offs $E_{p,cut}=300$~TeV for the NGC~4151 fit and $E_{p,cut}=70$~TeV for the NGC~1068 fit.

\section{Correlated neutrino -- hard X-ray variability?}

The hard X-ray emission from Seyfert galaxies is known to be violently variable \cite{2010MNRAS.408.1851L}.  This variability is naturally expected in the scenario of the corona heating by dissipation of magnetic field energy of the accretion disk. Similarly to the Solar corona, this process is driven by instabilities on different scales, and the largest scale instabilities, providing rare, but powerful energy releases, appear as flares. Following the logic of the previous section, one should also expect that the hard X-ray flaring activity in Seyfert galaxies should be accompanied by variations of the neutrino flux. The hard X-ray flux is expected to strongly correlate with the (sub-)GeV band neutrino flux, where most of the neutrino power is released. A more subtle relation between the hard X-ray and TeV-PeV neutrino flux may be expected because of possible variations of the slope and cut-off energy of the neutrino spectrum. 

To search for possible variability of neutrino flux, we used the PSlab data analysis package\footnote{https://github.com/icecube/PSLab\_PS\_analysis} that allows to perform a time-dependent unbinned-likelihood analysis~\cite{Braun:2009wp} of publicly available IceCube ten-year exposure dataset~\cite{IceCube:2019cia}. This is enabled via inclusion of the pre-defined time dependence of the probability density function for neutrino events, in the form of a Gaussian centered at a time moment $T_0$ and of variable width $\Delta T$. We perform a time-dependent analysis for the set of three sources selected in~\cite{Neronov:2023aks}: NGC 1068, NGC 4151 and NGC 3079. 

\begin{figure}
    \includegraphics[width=\linewidth]{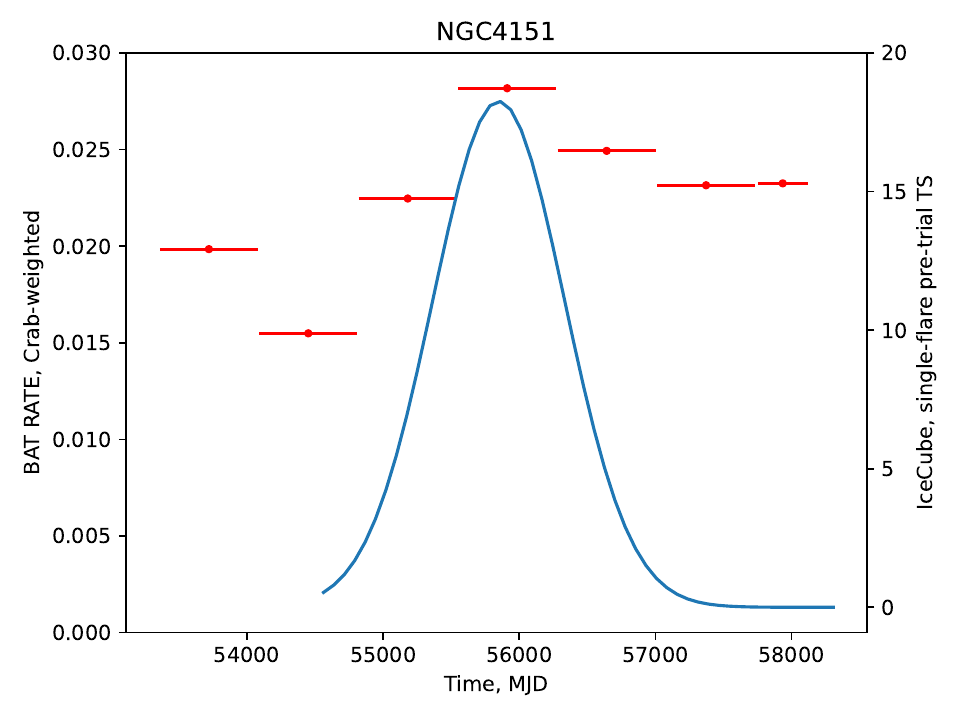}
 \includegraphics[width=\linewidth]{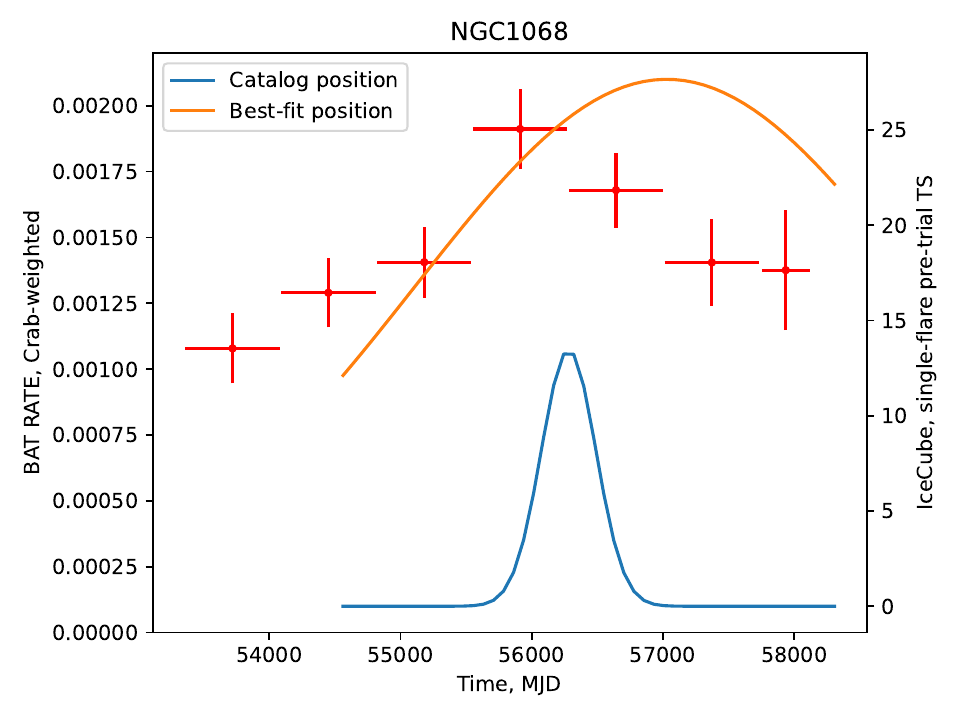}
 \caption{Red data points show SWIFT/BAT (15-195 keV) from 157-month catalog (https://swift.gsfc.nasa.gov/results/bs157mon/), compared to localisation of the IceCube signal in the time-dependent analysis with PSlab for NGC 4151 (top) and NGC 1068 (bottom), shown by  curves. In the bottom panel, blue  lower curve corresponds to the analysis at the catalog source position, the upper curve is for the best-fit neutrino source position.  
 }
    \label{fig:bat-vs-nu}
\end{figure}

Fig.~\ref{fig:bat-vs-nu} shows a comparison of the 15-195 keV band SWIFT/BAT lightcurve of NGC 4151 from the 157 month SWIFT/BAT survey\footnote{https://swift.gsfc.nasa.gov/results/bs157mon/} with the output of the PSLab analysis. There is a variability of the hard X-ray flux by a factor of up to 3  on the monthly time scale for NGC 4151, which is one of the brightest AGN on the sky in the hard X-ray band. Averaging the flux to two-year-long time bins highlights multi-year flux variations that are potentially detectable with neutrino telescope(s). During IceCube observations, the highest source flux was observed in the two-year time interval around 2012.

The time-dependent analysis of NGC 4151 with PSLab yields the best-fit time of the enhanced source flux period centered on $T_0=55849\pm194$ with the width $\Delta T=485$ containing $n_s=29\pm10$ counts with the test statistic $\mathrm{TS}=18.26$. Assuming the Wilks' theorem, this corresponds to the local significance at the level of $\sim3.2\sigma$, consistent with the results reported in Refs. \cite{IceCube:2022der,Neronov:2023aks,Abbasi:2024hwv}. Inclusion of the Gaussian time profile in the likelihood analysis allows us to localise the time interval which provides larger contribution to the signal significance. Remarkably, this time interval corresponds to the higher hard X-ray flux from the source, as can be seen from the top panel of Fig~\ref{fig:bat-vs-nu}.

The PSlab analysis for NGC 1068 provides the best-fit localisation of the most significant signal accumulation period $T_0=56281\pm68$ with width $\Delta T = 208$ and $n_s=23\pm8$ at the level of $\mathrm{TS}=13.46$, consistent with the results of Ref.~\cite{2021ApJ...920L..45A}. 
The detection significance is lower than that reported in Refs.  \cite{IceCube:2022der,Neronov:2023aks,Abbasi:2024hwv}, because the best-fit position of the source is slightly displaced from the direction toward NGC 1068. Performing the analysis for the  best-fit position from Ref.~\cite{Neronov:2023aks} one finds $T_0=57028\pm983$  and $\Delta T=1920$ with the Test Statistics value $TS=27.66$, consistent with the results of Refs. \cite{IceCube:2022der,Neronov:2023aks,Abbasi:2024hwv}. The bottom panel of Fig. \ref{fig:bat-vs-nu} shows a comparison of the results of the IceCube analysis with the SWIFT/BAT lightcurve of NGC 1068. The source is heavily absorbed in the hard X-ray band, so the detection significance is much lower, compared to NGC 4151, and monthly variability is not detectable. Variability on the two-year time scale is marginally observable, with a possible flux enhancement in 2012. Also in this case, the PSLab analysis indicates a localisaiton of the period with more important contribution to the signal-to-noise ratio around this period (for the analysis at the catalog NGC 1068 position) or later (for the analysis at the best-fit source position). 

NGC 3079 is also heavily absorbed in the hard X-ray band and its intrinsic variability is not detectable with SWIFT/BAT neither on monthly nor on yearly time scales. PSlab analysis for this source also finds the width of the activity period $\Delta T$ saturating the hard upper bound imposed in the analysis. 

In spite of the hints of faster accumulation of neutrino signal-to-noise ratio during the periods of higher hard X-ray flux, it is clear that the limited sensitivity of IceCube is insufficient for the detection of neutrino flux variations. A larger neutrino telescope is needed for clarification of the presence or absence of the neutrino -- hard X-ray correlated variability. 

%

\section{Contribution of Seyfert galaxies to the Astrophysical neutrino background.}

\begin{figure}
\includegraphics[width=\linewidth]{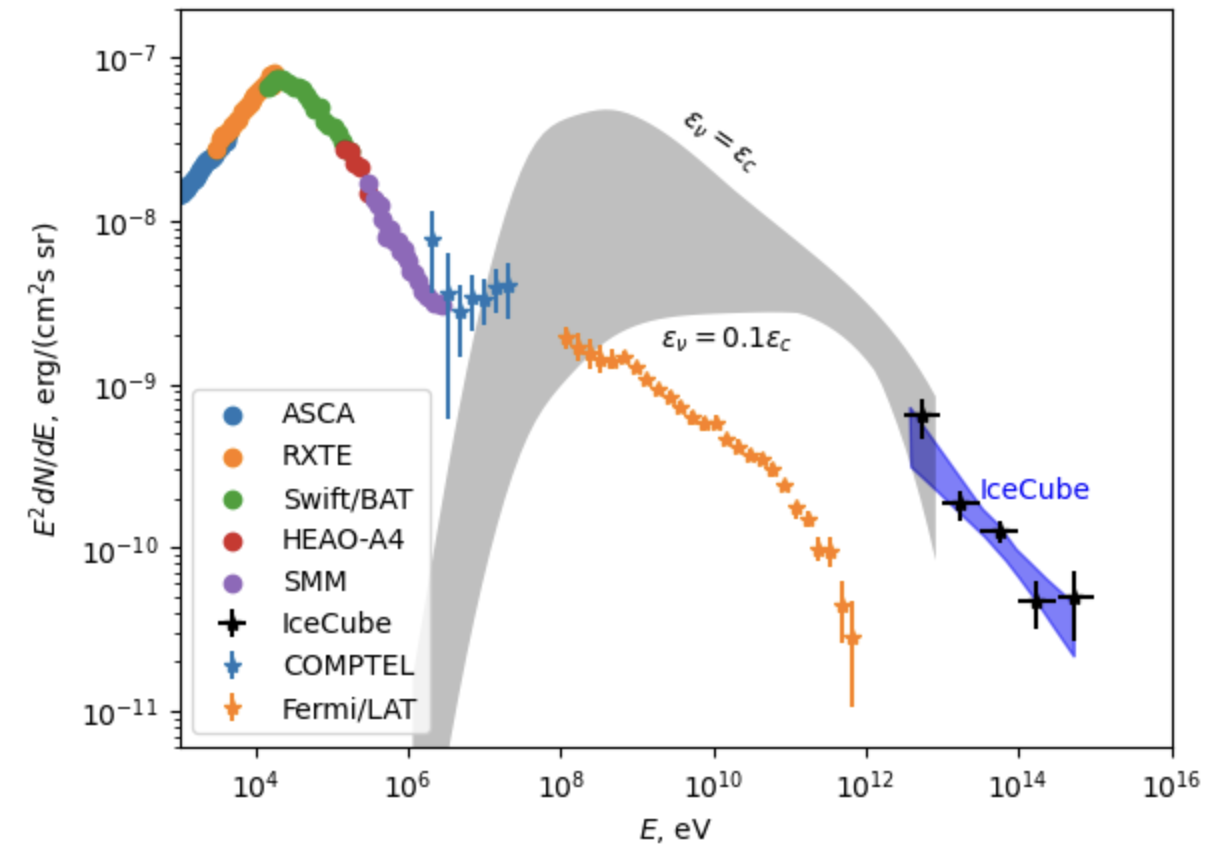}
    \caption{Spectra of diffuse photon \cite{Fermi-LAT:2014ryh,Neronov:2018ibl} and neutrino \cite{Abbasi:2024jro} backgrounds measured by a range of telescopes (specified in the legend). Gray shading shows possible contribution from Seyfert galaxies, calculated using either NGC 4151 or NGC 1068 template spectrum for the individual Seyfert galaxy source spectrum and $\epsilon_\nu$ between $0.1\epsilon_c$ and  $\epsilon_c$.}
    \label{fig:diffuse_spectrum}
\end{figure}

Hard X-ray emission from hot coronae of Seyfert galaxies is known to be the dominant source of the hard X-ray background. A possible common origin of corona heating and proton acceleration through dissipation of the magnetic energy of the disk suggests a relation between the overall levels of the X-ray ($E_{hX} F_{hX}$) and neutrino ($E_\nu F_\nu$) background energy fluxes, $E_\nu F_\nu\sim (\epsilon_\nu/\epsilon_c) E_XF_X$.
The neutrino flux is mostly concentrated in the GeV energy range, just above the energy threshold of the pion production reaction. Thus, the GeV neutrino background flux can be as high as reach $\sim 10^{-7}(\epsilon_\nu/\epsilon_c)$~erg/(cm$^2$s sr) in this band (see Fig. \ref{fig:diffuse_spectrum}).

The estimate of the level of neutrino background from Seyfert galaxies in the multi-TeV band is rather uncertain because of uncertainty of the spectral parameters for individual sources and of the distribution of these parameters across the population of Seyfert galaxies. For example, the fits to the neutrino spectra of NGC 4151 and NGC 1068 for the case $\epsilon_\nu\sim \epsilon_c$ shown in Fig. \ref{fig:spectrum} have the same slope of the proton spectrum, $\Gamma=2.4$, but different cut-off energies (500~TeV and 50 TeV). The distribution of cut-off energies of the proton spectra for different Seyfert galaxies is not known, and hence no reliable calculation of the predicted neutrino background from Seyfert galaxy population is possible.

Nevertheless, a rough estimate of the contribution of Seyfert galaxies to the astrophysical neutrino flux can be done assuming that the NGC1068 and NGC 4151 neutrino spectra are representative of the entire population and can be used as a template for the calculation of the overall flux from Seyfert galaxies.  Assuming that the neutrino and hard X-ray luminosities of individual sources are related as 
$
L_{\nu}\sim  (\epsilon_\nu/\epsilon_c) L_{hX}$
the calculation of the diffuse neutrino flux from Seyfert galaxies becomes identical to the calculation of the diffuse X-ray background produced by these sources \cite{Comastri:2004nd,Ueda:2014tma,Ambrosone:2024zrf}. The X-ray data constrain the the redshift evolution of the number density of Seyfert galaxies with given X-ray (and hence, neutrino) luminosity 
\begin{equation}
\frac{d\Phi (L_\nu,z)}{d\log(L_{\nu})dV}\sim \frac{\epsilon_\nu}{\epsilon_c}
\frac{d\Phi (L_{hX},z)}{d\log(L_{hX})dV}
\end{equation}
Repeating the calculation of the  the  X-ray background of Ref. \cite{Ueda:2014tma} for the neutrino case results in the diffuse neutrino background estimate shown in Fig. \ref{fig:diffuse_spectrum}. The gray shaded range in this figure includes a range of possible choices for the "typical" high-energy cut-off of the neutrino spectra (between 50~TeV inferred from the model of NGC 1068 and 500~TeV found in the model of NGC 4151). A similar uncertainty is also present in the slope of the neutrino spectrum (not shown in the figure). A qualitative conclusion that can be drawn from the figure is that it is possible that Seyfert galaxies provide an order-of-one contribution to the astrophysical neutrino flux at 10~TeV  \cite{Inoue:2019fil,Abbasi:2024hwv,Padovani:2024tgx}.

\section{Conclusions}

We have shown that the existing electromagnetic and neutrino data are consistent with a model of the non-thermal activity of the AGN central engine in which high-energy proton acceleration via turbulence and reconnection of magnetic field in the surface layer of the AGN accretion disk leads to both neutrino emission and heating of the disk surface and corona above it. In such a model, the hard X-ray and neutrino luminosity of Seyfert galaxies are naturally related. The hard X-ray brightest sources are expected to be the strongest neutrino emitters, and hard X-ray flux variations of individual sources are expected to be accompanied by variations of the neutrino flux. We have attempted a search for such correlated variations. Although a combination of IceCube and SWIFT/BAT data is consistent with a possibility of the correlated hard X-ray and neutrino activity, the sensitivity of IceCube is not sufficient for localization of the time periods of enhanced neutrino flux.  The model of common origin of the neutrino and hard X-ray emission also suggests that the cosmic neutrino background produced by Seyfert galaxies may be strongest in the GeV band and its high-energy tail can provide a sizable contribution to the astrophysical neutrino flux in multi-TeV range. 

\section{Acknowledgements}

We would like to thank M.Lemoine and F.Oikonomu for encouraging discussions. 

Work of M. Poleshchuk was supported by the National Research Foundation of Ukraine under project No. 2023.03/0149.

\bibliography{refs.bib}
\end{document}